\documentclass[aps,prb,superscriptaddress,reprint]{revtex4-1}%
\usepackage{epsfig,pslatex,latexsym,times,amssymb,amsmath,graphicx}
\usepackage{bm, float, lipsum}
\usepackage{amsmath}
\usepackage{amsfonts}
\usepackage{amssymb}
\usepackage{mathrsfs}
\usepackage{color}
\usepackage{graphicx}%
\providecommand{\U}[1]{\protect\rule{.1in}{.1in}}
\begin{document}
\author{N. J. Harmon}
\email{nicholas-harmon@uiowa.edu} 
\affiliation{Department of Physics and Astronomy and Optical Science and Technology Center, University of Iowa, Iowa City, Iowa
52242, USA}
\author{F. Maci\`a}
\affiliation{Department of Physics, New York University, New York, New York
10003, USA}
\author{F. Wang}
\affiliation{Department of Physics and Astronomy and Optical Science and Technology Center, University of Iowa, Iowa City, Iowa
52242, USA}
\author{M. Wohlgennant}
\affiliation{Department of Physics and Astronomy and Optical Science and Technology Center, University of Iowa, Iowa City, Iowa
52242, USA}
\author{A. D. Kent}
\affiliation{Department of Physics, New York University, New York, New York
10003, USA}
\author{M. E. Flatt\'e}
\affiliation{Department of Physics and Astronomy and Optical Science and Technology Center, University of Iowa, Iowa City, Iowa
52242, USA}
\date{\today}
\begin{abstract}
Random hyperfine fields are essential to mechanisms of low-field magnetoresistance in organic semiconductors. Recent experiments have shown that another type of random field --- fringe fields due to a nearby ferromagnet --- can also dramatically affect the magnetoresistance. A theoretical analysis of the effect of these fringe fields is challenging, as the fringe field magnitudes and their correlation lengths are orders of magnitude larger than that of the hyperfine couplings. We extend a recent theory of organic magnetoresistance to calculate the magnetoresistance with both hyperfine and fringe fields present. This theory describes several key features of the experimental fringe-field magnetoresistance, including the applied fields where the magnetoresistance reaches extrema, the applied field range of large magnetoresistance effects from the fringe fields, and the sign of the effect.
\end{abstract}
\title{Including fringe fields from a nearby ferromagnet in a percolation theory of organic magnetoresistance}
\maketitle

\emph{Introduction} - A major thrust of research in organic spintronics concerns transport through organic semiconductors sandwiched between either magnetic\cite{Dediu2002, Xiong2004} or non-magnetic\cite{Francis2004, Mermer2005b} electrodes. Both classes of devices display large magnetoresistance (MR), and have spurred significant experimental\cite{Kalinowski2003, Francis2004, Mermer2005b, Prigodin2006, Desai2007, Bloom2007} and theoretical\cite{Bobbert2007, Bobbert2009, Harmon2012a, Harmon2013a} interest. A complete description of the physics involved in this {\it organic magnetoresistance} (OMAR) is still evolving\cite{Wagemans2010}.
For the non-magnetic case, one explanation of the MR relies on the existence of random, uncorrelated hyperfine fields (HFs) that vary from one localizing center to another (referred to as HF-OMAR).\cite{Bobbert2007}
A recent experiment\cite{Wang2012} has tested the importance and influence of randomizing fields by situating a single \emph{electrically isolated} ferromagnet some specified distance from the organic semiconductor. When the ferromagnet is magnetically unsaturated, spatially varying fringe fields (FFs) emanate outside the ferromagnet, and these  FFs dramatically alter the MR lineshape (referred to as FF-OMAR).
Such FFs could also be important in experiments with magnetic electrodes, such as organic spin valves\cite{Nguyen2012}.
A theoretical explanation of the FF influence has proven difficult since the spatial distribution of FF's differs greatly from that of HF's;  the magnitude and correlation lengths are both more than one order of magnitude larger.

In this Rapid Communication, we demonstrate that a recently proposed theory of OMAR\cite{Harmon2012a, Harmon2012b, Harmon2012c}, based on percolation theory, offers a solution to the anomalous OMAR observed\cite{Wang2012} in the presence of FFs.
The main ingredient is that field-dependent spin transitions (due to FFs and HFs) open otherwise blocked hopping pathways which in turn alter the sample's resistance. We  find that a careful study of the statistics of the FFs identifies several regions of behavior, including a region close to the ferromagnetic film where FF \emph{gradients} govern the MR line shape, and a region farther away where HF re-emerge as an important element and where the dominant effect of the ferromagnet is through the FF magnitude. For regions closer to the ferromagnetic film, where FF gradients are important, we predict the size of the FF effect will not be very sensitive to the thickness of the ferromagnetic layer; as both the lateral size of the domains (and thus the field correlation length) and the magnetic moments will increase with thickness, the gradient should remain nearly constant.

\begin{figure}[ptbh]
 \begin{centering}
        \includegraphics[scale = 0.35,trim =1 5 10 10, angle = -0,clip]{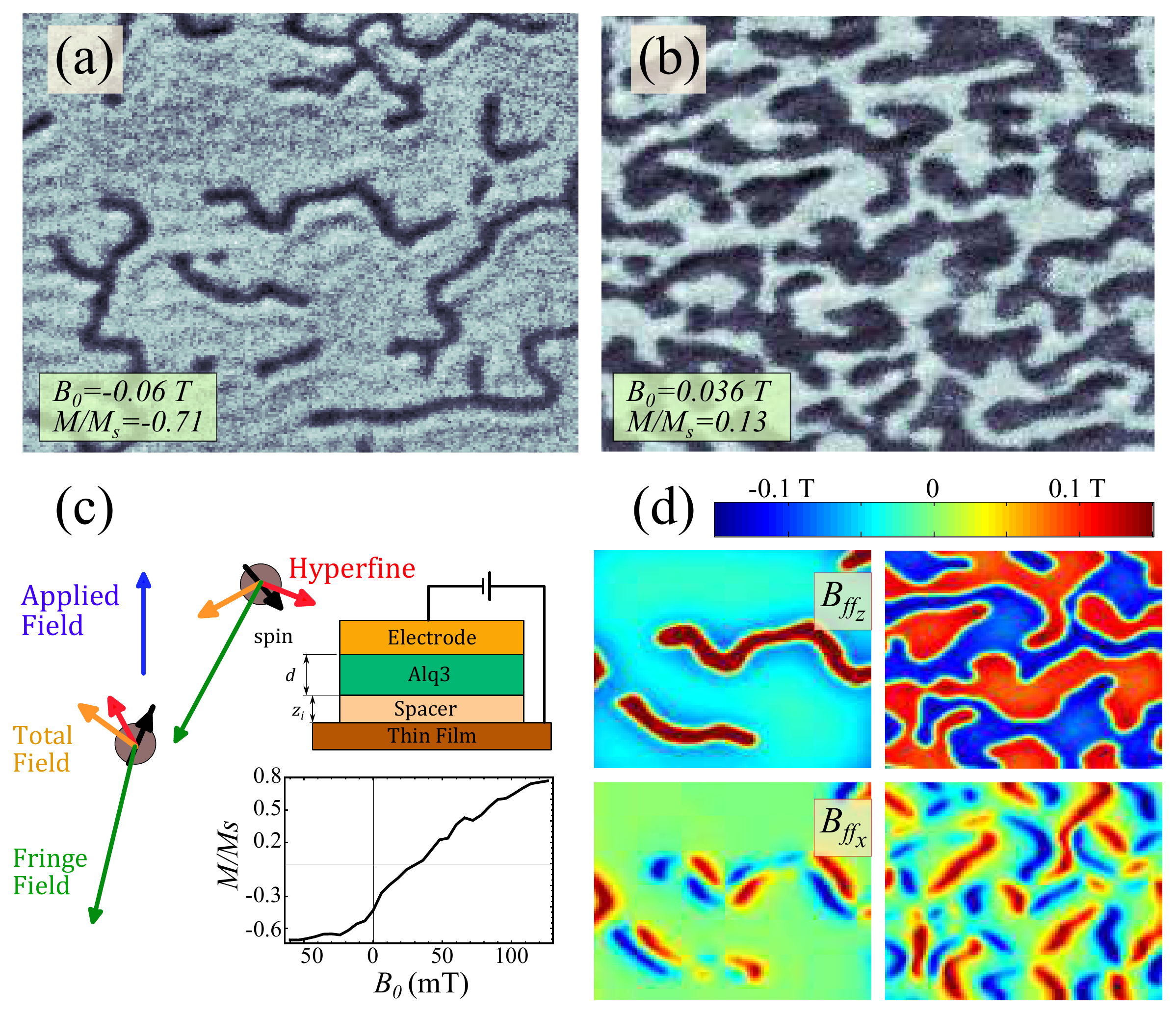}
        \caption[]
{(Color online) (a,b) Two examples of the ferromagnet with different domain configurations and normalized magnetizations ($M/M_s$) obtained by X-ray microscopy imaging in Ref.~\onlinecite{Wang2012}. (c) cartoon picture of two occupied sites and their local fields. Insets: experimental set-up and $M/M_s$ versus applied field $B_0$. (d) Fringe fields, $B_{ff_z}$ and $B_{ff_x}$, in a 2 $\mu$m x 2 $\mu$m area calculated 15 nm above the ferromagnet. }\label{fig:ffCartoon}
        \end{centering}
\end{figure}

\begin{figure}[ptbh]
 \begin{centering}
        \includegraphics[scale = 0.4,trim =50 233 260 5, angle = -0,clip]{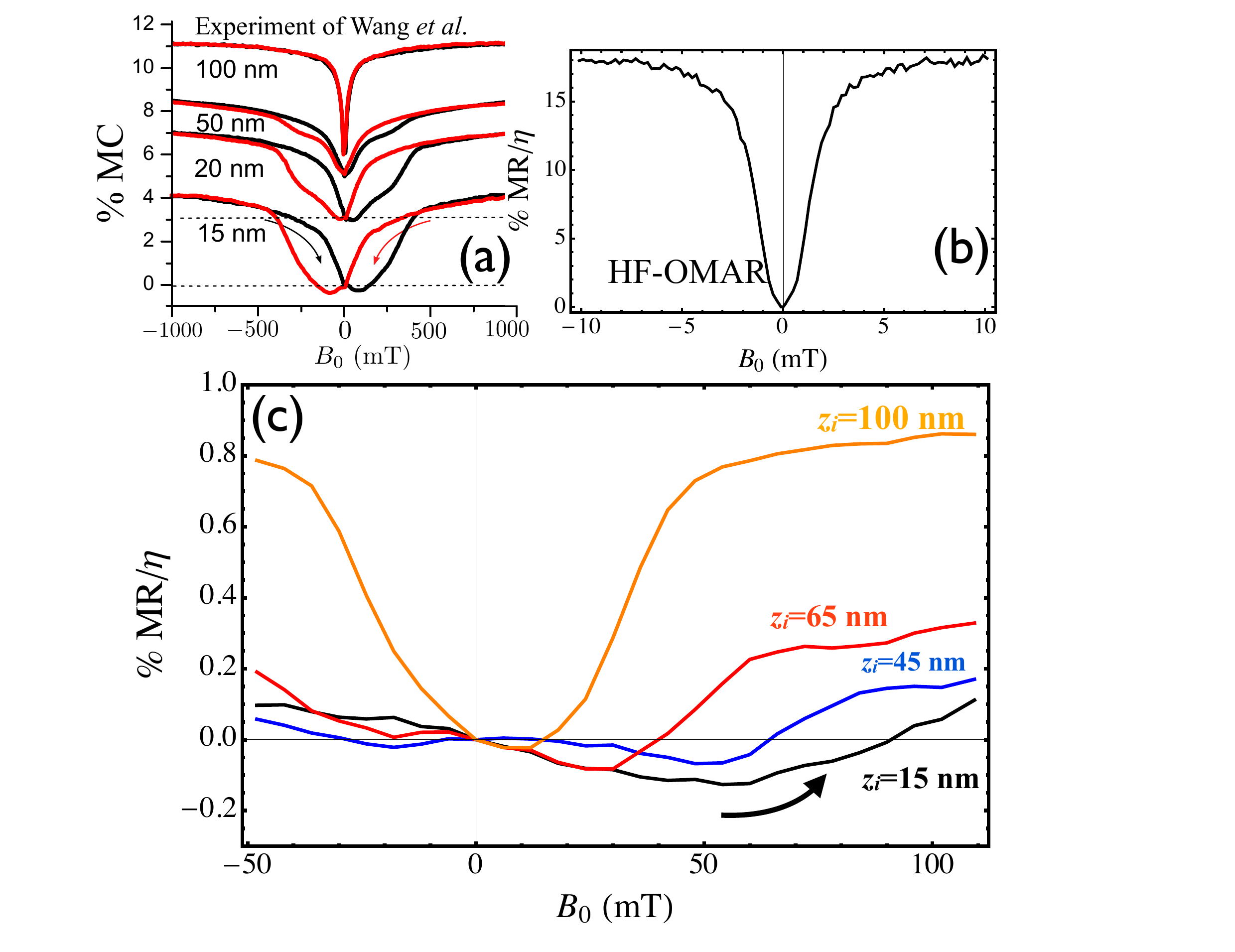}
        \caption[]
{(a) MC as measured for different spacer distances by Ref. \onlinecite{Wang2012}. 
(b) HF-OMAR calculated from our theory. 
(c) FF-OMAR calculated from our theory for several spacer distances with $a = 3$ mT and $d = 10$ nm. A moving average is taken to smooth the curves. The applied field range is limited by the availability of X-ray images at larger $B_0$. Thick arrow indicates the direction of field ramping.}\label{fig:ffMR}
        \end{centering}
\end{figure}

We now give a short description and summary of the experiments on FF-OMAR
reported in Ref. \onlinecite{Wang2012}. The organic semiconductor device, called a semi-spin valve (Figure \ref{fig:ffCartoon} (c)), consists
of a ferromagnetic Co/Pt multilayered film with perpendicular magnetic anisotropy followed by a
nonmagnetic metal, a bottom electrode, an organic layer, and a Ca top electrode. FF-OMAR is observed even
when the ferromagnetic layer is excluded from the current path, and therefore excludes effects such as spin-injection or tunneling anisotropic magnetoresistance.\cite{Grunewald2011}
The reversal of magnetization, $M$, in the ferromagnet occurs through nucleation, growth, and annihilation of magnetic domains.\cite{Hellwig2007}   
The saturation magnetization, $M_s$, is 5.4 $\times 10^5$ A/m. The properties of the ferromagnetic films have been further characterized in Ref. \onlinecite{Wang2012}.
The organic layer is a 30-
nm-thick film of tris(8-hydroxyquinolinealuminum (Alq$_3$). 
The bottom electrode also serves as a spacer layer of variable thickness, $z_i$, to separate the magnetoresistive
material, Alq$_3$, from the magnetic layer that is the source of the magnetic FFs.
The FF's strength and spatial-correlation length depend on the magnetic domain configuration as well as on the distance from the ferromagnet to the organic film. Typical experimental data is
shown in Figure \ref{fig:ffMR} (a). 
Unlike ordinary HF-OMAR, the FF-OMAR is hysteretic
and extends to a much larger field-scale. In Ref.~\onlinecite{Wang2012} transmission x-ray microscopy (TXM) based on the
x-ray magnetic circular dichroism (XMCD) effect was used to determine the Co/Pt layer's
microscopic magnetic domain structure as a function of the applied perpendicular field (see Figure \ref{fig:ffCartoon} (a,b)). 
We obtain here, through methods described below, the FF distribution at a given distance above the ferromagnetic film from magnetostatic
modeling (see Figure \ref{fig:ffCartoon} (d)).

\emph{Theory} - 
In disordered organic films transport occurs by hopping between localizing sites.
The fermionic nature of the charge carriers forbids the formation of doubly occupied sites in a triplet state. Hence a hopping polaron experiences a reduction in the number of accessible sites which affects the transport properties of the organic semiconductor. This leads to the formation of bottlenecks where a transport pathway is restricted due to the spin-blocking described. The situation can be alleviated when two polarons on nearby sites (a polaron pair) experience local magnetic fields which allow the charge-blocking triplet polaron pair to undergo a transition to the singlet state which allows for a doubly occupied site to form. However if the spin transition is slow enough, it may be more expedient for the blocked charge to bypass the other occupied site. 
This produces a competition between  two processes: formation of a doubly occupied site (e.g. a bipolaron or a doubly occupied deep trap) and disassociation of the polaron pair.  For local fields which are due to the nuclear spins, theory predicts the cross-over between the two processes can be seen in the MR when the hopping rate, $v_0$, is varied from slow hopping to fast hopping (compared to the strength of the HF).\cite{Schellekens2011, Harmon2012b} 

Solution to the spin-independent transport problem in such systems comes from percolation theory.\cite{Shklovskii1984}
The percolation theory solution of OMAR is encapsulated in the following equation for the threshold hopping distance, $r_c$,\cite{Harmon2012a, Harmon2012b, Harmon2012c}
\begin{equation}\label{eq:bondingCriterion1}
\int_0^{r_c}4 \pi  N_{eff} r^2 dr = B_c 
\end{equation}
with $N_{eff} = N  - N_T + \alpha p_{T\rightarrow S} N_T$ where $N_T = \frac{3}{4} N_1$ is the density of polarons that form a triplet spin state with some specific polaron; $N_1$ is the concentration of injected polarons with localization length $\ell$. $N$ is the density of sites in the system and considered to be $\gg N_1$ (i.e. dilute carrier concentration approximation). $\alpha$ is a number less than unity. Eq. (\ref{eq:bondingCriterion1}) is spin-dependent through the probability for a triplet polaron pair to transition to a singlet polaron pair, $p_{T\rightarrow S}$. $B_c$ is the average number of sites within a distance $r_c$ of one another that exist in the percolating cluster; in three dimensions, $B_c \approx 2.7$ as found from numerical simulations.\cite{Shklovskii1984}
The theory is valid for unipolar transport, though many of its qualitative features can also be applied to the bipolar regime\cite{Harmon2012b, Hu2007} probably relevant for the FF-OMAR experiments\cite{Wang2012}. 
The theory has been further bolstered by success in explaining features of experiments\cite{Wagemans2011}  in the bipolar regime.

After defining MR$\equiv \langle [R_c(B) - R_c(0)]/R_c(0) \rangle$ with $R_c = R_0 e^{2 r_c/\ell}$, we use Eq. (\ref{eq:bondingCriterion1})
to write\cite{Harmon2012a, Harmon2012b, Harmon2012c}
\begin{equation}\label{eq:fullMR}
\textrm{MR}
\approx \frac{\eta }{6 \pi y_{c}^2}\int_0^{2 \pi} \int_0^{\pi}  \int_0^{y_{c}}   \big\langle p_{S}(B)  - p_{S}(0) \big\rangle y^2 \sin\theta  dy d\theta d\phi,
\end{equation}
where $y = r/\ell$ and we have made the substitution $p_{T\rightarrow S} = \frac{1}{3} [1 - p_{S}]$ in terms of the easier to evaluate quantity $p_{S\rightarrow S} \equiv  p_S$ and $\eta = N_T/N$.\cite{Werner1977} 
Estimating $\eta$ is difficult since the carrier concentration is not known with precision and deeply charged traps likely exist that do not contribute to the carrier population but can contribute to the MR.\cite{Harmon2012c, Rybicki2012}
It has been established that\cite{Timmel1998}
 \begin{equation}\label{eq:fullPs}
p_{S} =
 \sum_{m, m'}| P_S^{mm'} |^2 \frac{ 1/\tau_h^2}{ (\omega_{m'} - \omega_{m})^2 + 1/\tau_h^2},
\end{equation}
where $P_S$ is the singlet projection operator, $m$ and $m'$ are indices denoting the eigenstates of the total Hamiltonian, $\omega$ represents the eigenvalues, and $\tau_h = v_0^{-1} \exp(2 r)$ is the hopping time between two sites separated by a distance $r$.
For concreteness, we choose $\ell = 0.2$ nm and $r_c = 1$ nm throughout this Rapid Communication.\cite{Gill1974,Martens2000}
We assume $v_0 = 10^2$ in frequency units of 1 mT.

The system Hamiltonian of the polaron pair, $\mathscr{H} = \mathscr{H}_0 + \mathscr{H}_{ff}$ is composed of terms that are responsible for `normal' OMAR,
\begin{equation}
\mathscr{H}_0 
 = 
 ( \bm{\omega}_{hf}(\bm{r}_1)+ \bm{\omega}_0) \cdot \bm{S}_1 +  ( \bm{\omega}_{hf}(\bm{r}_2)+ \bm{\omega}_0) \cdot \bm{S}_2,
 \end{equation}
 and a new term arising from the FFs:
 \begin{equation}
\mathscr{H}_{ff} = 
 \bm{\omega}_{ff}(\bm{r}_1) \cdot \bm{S}_1 +   \bm{\omega}_{ff}(\bm{r}_2) \cdot \bm{S}_2.
 \end{equation}
Both HFs and FFs have spatial dependence; however as shown below, the dependences are very different.
HFs are uncorrelated from site to site (on the order of 1 nm) and their magnitudes follow a Gaussian distribution of width $a$ which is on the order of a few milliteslas.
FFs are correlated over a larger distance ($\sim$ 100 nm) and their magnitudes are $\sim 100$ mT for a spacer length of $z_i = 15$ nm.\cite{Wang2012}

\emph{Calculation} - 
The existence of fringe fields adds considerable complexity to a calculation of Eq. (\ref{eq:fullMR}) since spatial variables must be accounted for in the eigensystem of $p_S$. Additionally the placement of a bottleneck could be in either a region of small or large FF in comparison to the HF; also the FF may either be constant or sharply vary between the two occupied sites. 
These issues are not encountered when the local fields are entirely uncorrelated (such as was the case with HF).
To account for these complexities, during the integration we sample pairs of sites (i.e. bottlenecks) randomly throughout the organic semiconductor volume. 
The spatial integrals (as well as the six integrals resulting from the average over hyperfine configurations) are done numerically by the Monte Carlo integration method:
\begin{eqnarray}\label{eq:P}
&& P(B_0) = \int_0^{2 \pi} \int_0^{\pi}  \int_0^{y_{c}} \langle p_{S} (y, \theta, \phi) \rangle y^2 \sin\theta  dy d\theta d\phi \approx{} \nonumber \\
 {}&&2 \pi  \cdot \pi \cdot y_{c} \frac{1}{K}\sum_{k=1}^{K} p_S(y_k, \theta_k, \phi_k) y_k^2 \sin\theta_k,
\end{eqnarray}
where the random numbers are chosen uniformly from the ranges $0 < y_k < y_{c}$, $0 < \theta_k < \pi$, and $0 < \phi_k < 2 \pi$. 
Angular brackets denote the averaging over hyperfine field distribution; the coordinates of the two hyperfine field vectors are also random variables included in the summation.
Convergence is slow; for the results presented herein, $K = 1-10$ million.
Whenever the applied field is changed, the calculation requires the FFs to be recalculated given the new domain configuration.
There are 31 X-ray images of the domains for 31 different applied fields within the magnetic switching regime. We calculate the MR at these applied fields in an 1 $\mu$m x 1 $\mu$m x $d$ volume. Larger volumes account for the domain structure more accurately but are computationally expensive. In the field-range where the ferromagnet is saturated (no FFs), we calculate MR from the HFs only.
\begin{figure}[ptbh]
 \begin{centering}
        \includegraphics[scale = 0.45,trim =210 52 300 250, angle = -0,clip]{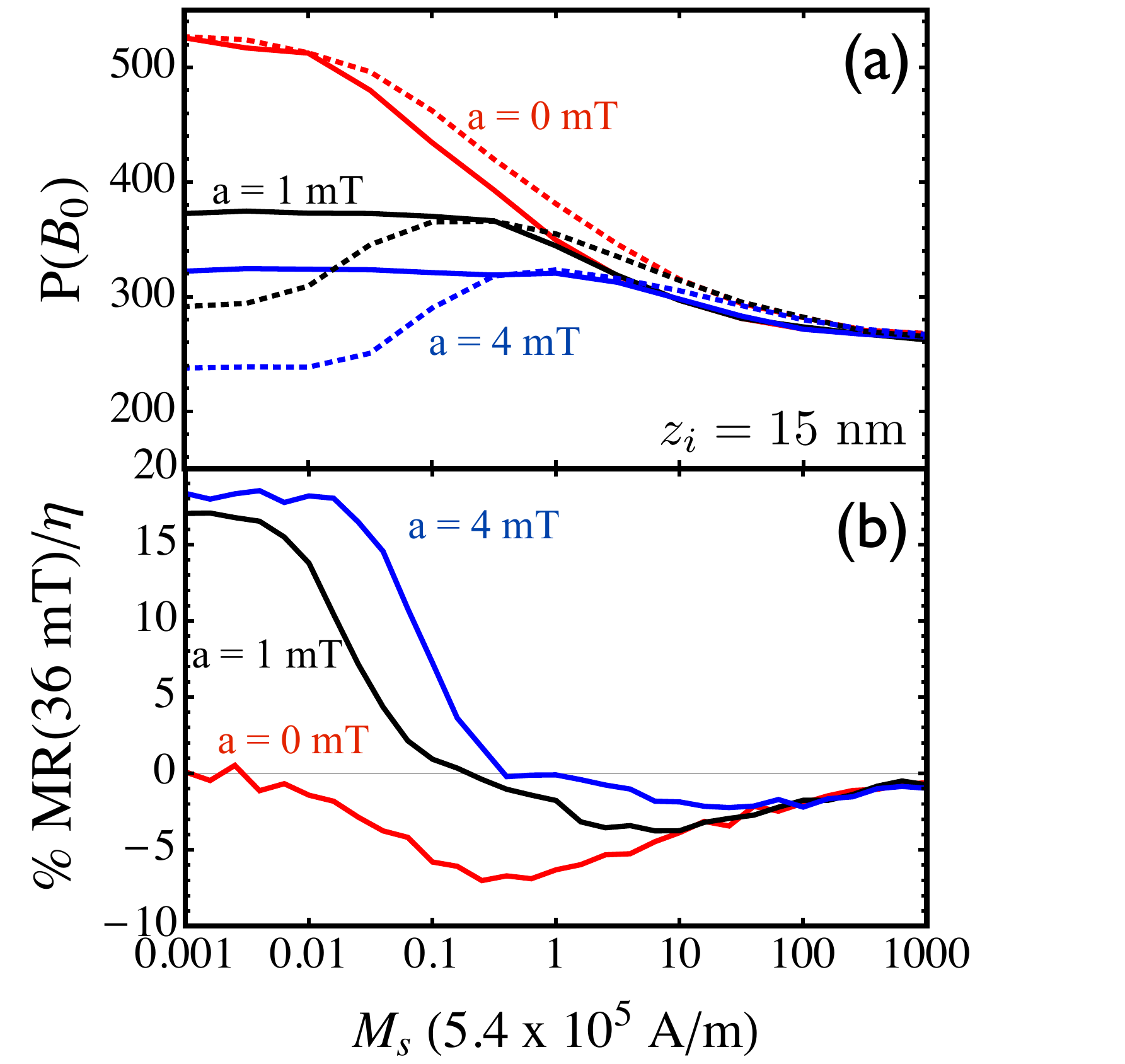}
        \caption[]
{(Color online) 
(a) The quantity $P(B_0)$ (Eq. \ref{eq:P}) at $B_0 = 0$ mT (dotted) and 36 mT (solid) for three different hyperfine couplings $a = 0$, 1, and 4 mT.
(b) The calculated \% MR for the same hyperfine couplings at 36 mT.
The field 36 mT is chosen since it nearly corresponds to the magnetization $M/M_s = 0$.
}\label{fig:MPlot}
        \end{centering}
\end{figure}
\begin{figure}[ptbh]
 \begin{centering}
        \includegraphics[scale = 0.38,trim =12 120 50 10, angle = -0,clip]{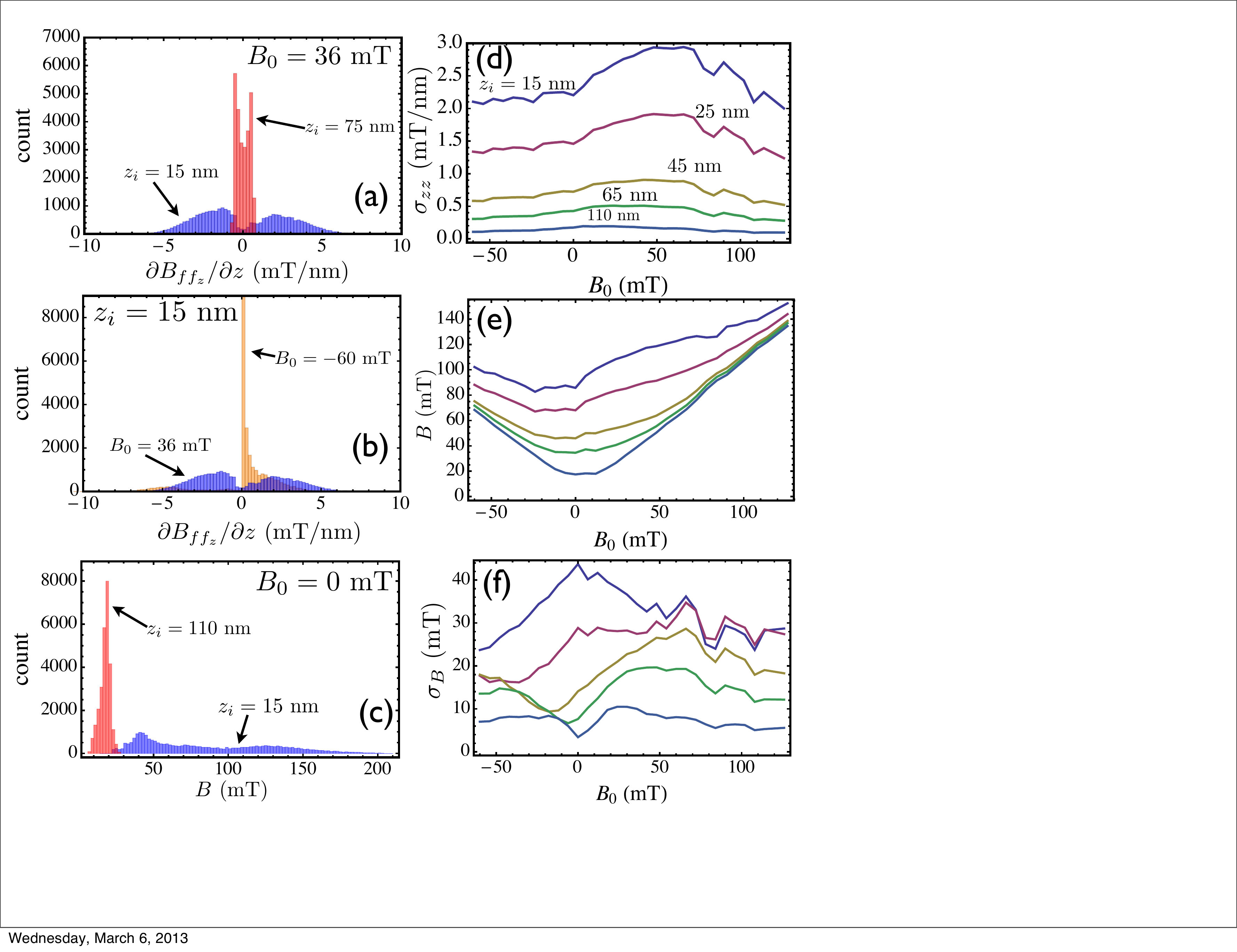}
        \caption[]
{(Color online) 
(a) and (b) are two examples of the partial derivative  of $B_{ff_z}$ with respect to $z$ distributions. 
(b) The width narrows as distance from the ferromagnet increases.
(b) The distributions are wider near $M = 0$ which corresponds to $B_0 \approx 36$ mT.
(c) An example of the total field (minus HF) distribution at $B_0 = 0$ mT for small and large $z_i$.
(d) Standard deviations, $\sigma_{zz}$, of $\partial_z b_z(\bm{r})$ computed at random points in the specified volume with bottom edge at $z_i$. 
Plots (e) and (f) follow the same direction of labels for different $z_i$.
 $\sigma_{ij}$ for $\{i,j\} \in \{x,y,z\}$) display similar behavior.
The average gradients (not shown), e.g. $\overline{\partial_z b_z(\bm{r})}$, do not stray significantly from zero.
(e) The average total field (minus HF) magnitude as a function of applied field for several different $z_i$.
(f) The standard deviation of the total field (minus HF) magnitude as a function of applied field for several different $z_i$.
The vector $\bm{B}$ is $\bm{B}_0 + \bm{B}_{ff}$.
}\label{fig:gradientFig}
        \end{centering}
\end{figure}

\emph{Results}-
Figure \ref{fig:ffMR}(c) shows our main results for the MR, to be compared with the experimental results from Ref.~\onlinecite{Wang2012}, Figure \ref{fig:ffMR}(a). Note that Ref.~\onlinecite{Wang2012}'s results are for the magnetoconductivity (MC), which implies that their MR is of opposite sign to our calculations (when a MR is calculated for unipolar transport, the MR seen in bipolar transport usually has opposite sign due to the fact that forming a doubly occupied site, i.e. an exciton, reduces current\cite{Hu2007, Nguyen2010b, Harmon2012b}).  The experimental and theoretical curves exhibit similar trends as the distance between the organic and ferromagnetic layer increases. For example, a region of negative MR is seen at small positive fields for small distances, and the area of this region decreases as the distance increases. For even larger distances eventually HF-OMAR should emerge as FFs become weaker. This is artificially demonstrated in Figure \ref{fig:MPlot} (b) by controlling the magnitude of the FFs by adjusting $M_s$. As the FFs (as well as the size of their gradients) are reduced, HF-OMAR is retained ($\sim 18.5$ \% in this case). By adjusting the HF coupling, it becomes apparent that there is a competition between FF and HFs in determining the overall behavior. 
The X-ray images exist for only ramping up the field so the hysteretic nature of the FF-OMAR cannot be seen from the calculation.

A natural question arises regarding the minima observed in the MR line shape.
 Why and where do the minima occur?
We find that this feature corresponds well to behavior observed in the FF statistics.
From the analysis of the FF statistics  performed in Ref. \onlinecite{Wang2012} it is known that the FF correlation length was two orders of magnitude larger than that of the HF correlation length ($\sim 100$ nm versus 1 nm). However, possibly a better indication of the efficacy of FFs in altering the MR is the FF gradient between the bottleneck sites.\cite{Cohen2009} We find that the gradients can be comparable to the HF gradients so it is reasonable to expect the FF to modulate the MR.

Our statistical analysis is depicted in Figure \ref{fig:gradientFig} where the averages and standard deviations of FFs are determined.
Several conclusions can be derived from the FF statistics.
One key point is that even at $B_0 = 0$ mT, there still exists large FFs such that $\overline{B_{ff}}\gg B_{hf}$. This is shown in Figure \ref{fig:gradientFig} (e) for the smaller $z_i$ where $\overline{B_{ff}} \approx 100$ mT at $B_0 = 0$ mT ($z_i = 15$ nm).
While the width of the distribution is $\approx 30$ mT, the FFs still quench practically all the HFs.
In view of Figure \ref{fig:gradientFig} (f) also, the situation is unchanged; the average total field increases as does the width but still all HFs are overpowered. 
This suggests that the mechanism that produces the deviation from the Lorentzian line shape --- referred to here as `ears' --- is \emph{not} due to a mixture of HF-OMAR components with uniform FFs adding to the applied field (this is the ``uncorrelated $B$-field model" of Ref. \onlinecite{Wang2012}).
The `ear' cannot be due to any decrease in net field; $\overline{B}$ increases while $\sigma_{B}$ decreases modestly (Figure \ref{fig:gradientFig} (e,f)) and an increase in total field tends to result in MR $>$ 0 which is not observed in that field range.
However if the distance $z_i$ increases, the situation changes because the average total field reduces dramatically, as demonstrated in Figure \ref{fig:gradientFig} (c). For such cases, the average field $\overline{B}$ in a small finite field can actually be larger than the average field at zero field; this effect reduces the resistance and shows up as MR $<$ 0 (not shown).

The statistical analysis of the previous paragraph points to the OMAR `ear' for low $z_i$ being due to the gradients in the FFs. 
Figure \ref{fig:gradientFig} (a) shows that typical sizes of the FF gradients can be up to a few millitesla per nanometer which would make the effect larger or at least comparable to HF-OMAR.
The trend in $\sigma_{zz}$ with applied field correlates with the `ear' which lend further credence to the importance of the FF gradients.
This is explainable from the fact that at $M = 0$, there are equal numbers of up and down domains which give rise to maximally varying FFs; in short, the probability that the two sites of the bottleneck feel local fields that are different is a maximum at $M  =0$. As the magnetization approaches saturation, one expects the standard deviation to be zero as the FFs are quenched - however the quality of the X-ray images diminishes in such cases and the domain configurations cannot be extracted for $M/M_s$ nearer to unity.
HF-OMAR does not immediately emerge when the FF gradients get smaller than the HF gradients - 65 nm curve in Figure \ref{fig:gradientFig} (d) is very narrow so FF gradients are small. However the MR line shape is still significantly modified from that of HF-OMAR (Figure \ref{fig:ffMR} (b)). While the gradients are small, the HFs still are influenced by the FF (even at zero applied field), as the FFs, though not varying spatially much, are still larger than the HFs.

Figure \ref{fig:MPlot} (a) shows the quantity $P(B_0)$ for three different hyperfine couplings, $a$.
First consider non-zero HF and the $B_0$ = 0 mT lines (dotted): 
for the smallest $M_s$, the FFs are negligible so $P$ and MR are independent of $M_s$. 
When $M_s$ increases to where $\overline{B} \sim a$, the FFs still have very small gradients (e.g. the $\sigma_{zz}$ of Figure \ref{fig:gradientFig} will scale linearly with $M_s$ so at $M_s = 0.01$, $\sigma_{zz} \approx 0.03$ mT/nm $\ll \sigma_{HF}\approx 1$ mT/nm) so the FFs effectively act as uniform fields across a site pair. This shows up as an increase in $P(0)$ and a decrease in MR(36 mT) since even though the applied field is fixed, increasing $M_s$ is effectively increasing the uniform field felt by a site pair.
Eventually $M_s$ can be increased enough to where $\sigma_{zz}$ becomes comparable to $\sigma_{HF}$; the variations in FFs between the site pair can now cause spin transitions more effectively. Since the FFs random variations are larger at 36 mT than at 0 mT, the MR is negative. For larger $a$, the $P(0)$ and $MR$ curve shift to higher $M_s$ since $\sigma_{HF}$ is larger.
When $a = 0$ mT, a different picture emerges; at $M_s = 0$, there can be no spin-transitions between singlets and triplets so $p_S = 1$ and $P(0) = \frac{4\pi}{3}y_c^3 \approx 523.6$. In the opposite limit of large $M_s$, $p_S = 1/2$ so $P(0)$ converges to $\approx  \frac{4\pi}{6}y_c^3$.

\emph{Conclusions}-
The present approach has shown that the salient features of FF-OMAR are explainable by our recent theory of OMAR based on calculating magneto-transport with percolation theory. This work serves as a springboard for further calculations to be performed within other OMAR theories - especially those that handle ambipolar carriers.\cite{Schellekens2011} 
To find better agreement with experiment requires studying the dependence of MR on the hopping rate, $v_0$.
That task necessitates more intensive numerical work and thus has been avoided in this current discussion.
This line of inquiry is also expected to elucidate how the FF-mechanism is similar to either the HF or the $\Delta g$ mechanisms.\cite{Wang2008}

We acknowledge support from an ARO MURI and stimulating discussions with P. A.  Bobbert.
F. M. acknowledges support from MC-IOF 253214.


\begin{thebibliography}{10}%
\makeatletter
\providecommand \@ifxundefined [1]{%
 \ifx #1\undefined \expandafter \@firstoftwo
 \else \expandafter \@secondoftwo
\fi
}%
\providecommand \@ifnum [1]{%
 \ifnum #1\expandafter \@firstoftwo
 \else \expandafter \@secondoftwo
\fi
}%
\providecommand \enquote [1]{``#1''}%
\providecommand \bibnamefont  [1]{#1}%
\providecommand \bibfnamefont [1]{#1}%
\providecommand \citenamefont [1]{#1}%
\providecommand\href[0]{\@sanitize\@href}%
\providecommand\@href[1]{\endgroup\@@startlink{#1}\endgroup\@@href}%
\providecommand\@@href[1]{#1\@@endlink}%
\providecommand \@sanitize [0]{\begingroup\catcode`\&12\catcode`\#12\relax}%
\@ifxundefined \pdfoutput {\@firstoftwo}{%
 \@ifnum{\z@=\pdfoutput}{\@firstoftwo}{\@secondoftwo}%
}{%
 \providecommand\@@startlink[1]{\leavevmode}%
 \providecommand\@@endlink[0]{}%
}{%
 \providecommand\@@startlink[1]{%
  \leavevmode
  \pdfstartlink
   attr{/Border[0 0 1 ]/H/I/C[0 1 1]}%
   user{/Subtype/Link/A<</Type/Action/S/URI/URI(#1)>>}%
  \relax
 }%
 \providecommand\@@endlink[0]{\pdfendlink}%
}%
\providecommand \url  [0]{\begingroup\@sanitize \@url }%
\providecommand \@url [1]{\endgroup\@href {#1}{\urlprefix}}%
\providecommand \urlprefix [0]{URL }%
\providecommand \Eprint[0]{\href }%
\@ifxundefined \urlstyle {%
  \providecommand \doi [1]{doi:\discretionary{}{}{}#1}%
}{%
  \providecommand \doi [0]{doi:\discretionary{}{}{}\begingroup
  \urlstyle{rm}\Url }%
}%
\providecommand \doibase [0]{http://dx.doi.org/}%
\providecommand \Doi[1]{\href{\doibase#1}}%
\providecommand \bibAnnote [3]{%
  \BibitemShut{#1}%
  \begin{quotation}\noindent
    \textsc{Key:}\ #2\\\textsc{Annotation:}\ #3%
  \end{quotation}%
}%
\providecommand \bibAnnoteFile [2]{%
  \IfFileExists{#2}{\bibAnnote {#1} {#2} {\input{#2}}}{}%
}%
\providecommand \typeout [0]{\immediate \write \m@ne }%
\providecommand \selectlanguage [0]{\@gobble}%
\providecommand \bibinfo [0]{\@secondoftwo}%
\providecommand \bibfield [0]{\@secondoftwo}%
\providecommand \translation [1]{[#1]}%
\providecommand \BibitemOpen[0]{}%
\providecommand \bibitemStop [0]{}%
\providecommand \bibitemNoStop [0]{.\EOS\space}%
\providecommand \EOS [0]{\spacefactor3000\relax}%
\providecommand \BibitemShut [1]{\csname bibitem#1\endcsname}%
\bibitem{Dediu2002}%
  \BibitemOpen
  \bibfield{author}{%
  \bibinfo {author} {\bibfnamefont{V.}~\bibnamefont{Dediu}}, \bibinfo {author}
  {\bibfnamefont{M.}~\bibnamefont{Murgia}}, \bibinfo {author}
  {\bibfnamefont{F.~C.}\ \bibnamefont{Matacotta}}, \bibinfo {author}
  {\bibfnamefont{C.}~\bibnamefont{Taliani}},\ and\ \bibinfo {author}
  {\bibfnamefont{S.}~\bibnamefont{Barbanera}},\ }%
  \bibfield{journal}{%
  \bibinfo {journal} {Solid State Communications}\ }%
  \textbf{\bibinfo {volume} {122}},\ \bibinfo {pages} {181} (\bibinfo {year}
  {2002})%
  \bibAnnoteFile{NoStop}{Dediu2002}%
\bibitem{Xiong2004}%
  \BibitemOpen
  \bibfield{author}{%
  \bibinfo {author} {\bibfnamefont{Z.~H.}\ \bibnamefont{Xiong}}, \bibinfo
  {author} {\bibfnamefont{D.}~\bibnamefont{Wu}}, \bibinfo {author}
  {\bibfnamefont{Z.~V.}\ \bibnamefont{Vardeny}},\ and\ \bibinfo {author}
  {\bibfnamefont{J.}~\bibnamefont{Shi}},\ }%
  \bibfield{journal}{%
  \bibinfo {journal} {Nature}\ }%
  \textbf{\bibinfo {volume} {427}},\ \bibinfo {pages} {821} (\bibinfo {year}
  {2004})%
  \bibAnnoteFile{NoStop}{Xiong2004}%
\bibitem{Francis2004}%
  \BibitemOpen
  \bibfield{author}{%
  \bibinfo {author} {\bibfnamefont{T.~L.}\ \bibnamefont{Francis}}, \bibinfo
  {author} {\bibfnamefont{O.}~\bibnamefont{Mermer}}, \bibinfo {author}
  {\bibfnamefont{G.}~\bibnamefont{Veeraraghavan}},\ and\ \bibinfo {author}
  {\bibfnamefont{M.}~\bibnamefont{Wohlgenannt}},\ }%
  \bibfield{journal}{%
  \bibinfo {journal} {New Journal of Physics}\ }%
  \textbf{\bibinfo {volume} {6}},\ \bibinfo {pages} {185} (\bibinfo {year}
  {2004})%
  \bibAnnoteFile{NoStop}{Francis2004}%
\bibitem{Mermer2005b}%
  \BibitemOpen
  \bibfield{author}{%
  \bibinfo {author} {\bibfnamefont{O.}~\bibnamefont{Mermer}}, \bibinfo {author}
  {\bibfnamefont{G.}~\bibnamefont{Veeraraghavan}}, \bibinfo {author}
  {\bibfnamefont{T.~L.}\ \bibnamefont{Francis}},\ and\ \bibinfo {author}
  {\bibfnamefont{M.}~\bibnamefont{Wohlgenannt}},\ }%
  \bibfield{journal}{%
  \bibinfo {journal} {Solid State Communications}\ }%
  \textbf{\bibinfo {volume} {134}},\ \bibinfo {pages} {631} (\bibinfo {year}
  {2005})%
  \bibAnnoteFile{NoStop}{Mermer2005b}%
\bibitem{Kalinowski2003}%
  \BibitemOpen
  \bibfield{author}{%
  \bibinfo {author} {\bibfnamefont{J.}~\bibnamefont{Kalinowski}}, \bibinfo
  {author} {\bibfnamefont{M.}~\bibnamefont{Cocchi}}, \bibinfo {author}
  {\bibfnamefont{D.}~\bibnamefont{Virgili}}, \bibinfo {author}
  {\bibfnamefont{P.~D.}\ \bibnamefont{Marco}},\ and\ \bibinfo {author}
  {\bibfnamefont{V.}~\bibnamefont{Fattori}},\ }%
  \bibfield{journal}{%
  \bibinfo {journal} {Chem. Phys. Lett.}\ }%
  \textbf{\bibinfo {volume} {380}},\ \bibinfo {pages} {710} (\bibinfo {year}
  {2003})%
  \bibAnnoteFile{NoStop}{Kalinowski2003}%
\bibitem{Prigodin2006}%
  \BibitemOpen
  \bibfield{author}{%
  \bibinfo {author} {\bibfnamefont{V.~N.}\ \bibnamefont{Prigodin}}, \bibinfo
  {author} {\bibfnamefont{J.~D.}\ \bibnamefont{Bergeson}}, \bibinfo {author}
  {\bibfnamefont{D.~M.}\ \bibnamefont{Lincoln}},\ and\ \bibinfo {author}
  {\bibfnamefont{A.~J.}\ \bibnamefont{Epstein}},\ }%
  \bibfield{journal}{%
  \bibinfo {journal} {Synthetic Metals}\ }%
  \textbf{\bibinfo {volume} {156}},\ \bibinfo {pages} {757} (\bibinfo {year}
  {2006})%
  \bibAnnoteFile{NoStop}{Prigodin2006}%
\bibitem{Desai2007}%
  \BibitemOpen
  \bibfield{author}{%
  \bibinfo {author} {\bibfnamefont{P.}~\bibnamefont{Desai}}, \bibinfo {author}
  {\bibfnamefont{P.}~\bibnamefont{Shakya}}, \bibinfo {author}
  {\bibfnamefont{T.}~\bibnamefont{Kreouzis}},\ and\ \bibinfo {author}
  {\bibfnamefont{W.~P.}\ \bibnamefont{Gillin}},\ }%
  \bibfield{journal}{%
  \bibinfo {journal} {Phys. Rev. B}\ }%
  \textbf{\bibinfo {volume} {76}},\ \bibinfo {pages} {235202} (\bibinfo {year}
  {2007})%
  \bibAnnoteFile{NoStop}{Desai2007}%
\bibitem{Bloom2007}%
  \BibitemOpen
  \bibfield{author}{%
  \bibinfo {author} {\bibfnamefont{F.~L.}\ \bibnamefont{Bloom}}, \bibinfo
  {author} {\bibfnamefont{W.}~\bibnamefont{Wagemans}}, \bibinfo {author}
  {\bibfnamefont{M.}~\bibnamefont{Kemerink}},\ and\ \bibinfo {author}
  {\bibfnamefont{B.}~\bibnamefont{Koopmans}},\ }%
  \bibfield{journal}{%
  \bibinfo {journal} {Phys. Rev. Lett.}\ }%
  \textbf{\bibinfo {volume} {99}},\ \bibinfo {pages} {257201} (\bibinfo {year}
  {2007})%
  \bibAnnoteFile{NoStop}{Bloom2007}%
\bibitem{Bobbert2007}%
  \BibitemOpen
  \bibfield{author}{%
  \bibinfo {author} {\bibfnamefont{P.~A.}\ \bibnamefont{Bobbert}}, \bibinfo
  {author} {\bibfnamefont{T.~D.}\ \bibnamefont{Nguyen}}, \bibinfo {author}
  {\bibfnamefont{F.~W.~A.}\ \bibnamefont{van Oost}}, \bibinfo {author}
  {\bibfnamefont{B.}~\bibnamefont{Koopmans}},\ and\ \bibinfo {author}
  {\bibfnamefont{M.}~\bibnamefont{Wohlgenannt}},\ }%
  \bibfield{journal}{%
  \bibinfo {journal} {Phys. Rev. Lett.}\ }%
  \textbf{\bibinfo {volume} {99}},\ \bibinfo {pages} {216801} (\bibinfo {year}
  {2007})%
  \bibAnnoteFile{NoStop}{Bobbert2007}%
\bibitem{Bobbert2009}%
  \BibitemOpen
  \bibfield{author}{%
  \bibinfo {author} {\bibfnamefont{P.~A.}\ \bibnamefont{Bobbert}}, \bibinfo
  {author} {\bibfnamefont{W.}~\bibnamefont{Wagemans}}, \bibinfo {author}
  {\bibfnamefont{F.~W.~A.}\ \bibnamefont{Oost}}, \bibinfo {author}
  {\bibfnamefont{B.}~\bibnamefont{Koopmans}},\ and\ \bibinfo {author}
  {\bibfnamefont{M.}~\bibnamefont{Wohlgenannt}},\ }%
  \bibfield{journal}{%
  \bibinfo {journal} {Phys. Rev. Lett.}\ }%
  \textbf{\bibinfo {volume} {102}},\ \bibinfo {pages} {156604} (\bibinfo {year}
  {2009})%
  \bibAnnoteFile{NoStop}{Bobbert2009}%
\bibitem{Harmon2012a}%
  \BibitemOpen
  \bibfield{author}{%
  \bibinfo {author} {\bibfnamefont{N.~J.}\ \bibnamefont{Harmon}}\ and\ \bibinfo
  {author} {\bibfnamefont{M.~E.}\ \bibnamefont{Flatt\'e}},\ }%
  \bibfield{journal}{%
  \bibinfo {journal} {Phys. Rev. Lett.}\ }%
  \textbf{\bibinfo {volume} {108}},\ \bibinfo {pages} {186602} (\bibinfo {year}
  {2012})%
  \bibAnnoteFile{NoStop}{Harmon2012a}%
\bibitem{Harmon2013a}%
  \BibitemOpen
  \bibfield{author}{%
  \bibinfo {author} {\bibfnamefont{N.~J.}\ \bibnamefont{Harmon}}\ and\ \bibinfo
  {author} {\bibfnamefont{M.~E.}\ \bibnamefont{Flatt\'e}},\ }%
  \bibinfo {journal} {submitted}%
  \bibAnnoteFile{NoStop}{Harmon2013a}%
\bibitem{Wagemans2010}%
  \BibitemOpen
\bibfield{journal}{%
    }%
  \bibfield{author}{%
  \bibinfo {author} {\bibfnamefont{W.}~\bibnamefont{Wagemans}}\ and\ \bibinfo
  {author} {\bibfnamefont{B.}~\bibnamefont{Koopmans}},\ }%
  \bibfield{journal}{%
  \bibinfo {journal} {Phys. Satus Solidi B}\ }%
  \textbf{\bibinfo {volume} {248}},\ \bibinfo {pages} {1029} (\bibinfo {year}
  {2011})%
  \bibAnnoteFile{NoStop}{Wagemans2010}%
\bibitem{Wang2012}%
  \BibitemOpen
  \bibfield{author}{%
  \bibinfo {author} {\bibfnamefont{F.}~\bibnamefont{Wang}}, \bibinfo {author}
  {\bibfnamefont{F.}~\bibnamefont{Maci\'a}}, \bibinfo {author}
  {\bibfnamefont{M.}~\bibnamefont{Wohlgenannt}}, \bibinfo {author}
  {\bibfnamefont{A.~D.}\ \bibnamefont{Kent}},\ and\ \bibinfo {author}
  {\bibfnamefont{M.~E.}\ \bibnamefont{Flatt\'e}},\ }%
  \bibfield{journal}{%
  \bibinfo {journal} {Phys. Rev. X}\ }%
  \textbf{\bibinfo {volume} {2}},\ \bibinfo {pages} {021013} (\bibinfo {year}
  {2012})%
  \bibAnnoteFile{NoStop}{Wang2012}%
\bibitem{Nguyen2012}%
  \BibitemOpen
  \bibfield{author}{%
  \bibinfo {author} {\bibfnamefont{T.~D.}\ \bibnamefont{Nguyen}}, \bibinfo
  {author} {\bibfnamefont{E.}~\bibnamefont{Ehrenfreund}},\ and\ \bibinfo
  {author} {\bibfnamefont{Z.~V.}\ \bibnamefont{Vardeny}},\ }%
  \bibfield{journal}{%
  \bibinfo {journal} {Science}\ }%
  \textbf{\bibinfo {volume} {337}},\ \bibinfo {pages} {204} (\bibinfo {year}
  {2012})%
  \bibAnnoteFile{NoStop}{Nguyen2012}%
\bibitem{Harmon2012b}%
  \BibitemOpen
  \bibfield{author}{%
  \bibinfo {author} {\bibfnamefont{N.~J.}\ \bibnamefont{Harmon}}\ and\ \bibinfo
  {author} {\bibfnamefont{M.~E.}\ \bibnamefont{Flatt\'e}},\ }%
  \bibfield{journal}{%
  \bibinfo {journal} {Phys. Rev. B}\ }%
  \textbf{\bibinfo {volume} {85}},\ \bibinfo {pages} {075204} (\bibinfo {year}
  {2012})%
  \bibAnnoteFile{NoStop}{Harmon2012b}%
\bibitem{Harmon2012c}%
  \BibitemOpen
  \bibfield{author}{%
  \bibinfo {author} {\bibfnamefont{N.~J.}\ \bibnamefont{Harmon}}\ and\ \bibinfo
  {author} {\bibfnamefont{M.~E.}\ \bibnamefont{Flatt\'e}},\ }%
  \bibfield{journal}{%
  \bibinfo {journal} {Phys. Rev. B}\ }%
  \textbf{\bibinfo {volume} {85}},\ \bibinfo {pages} {245213} (\bibinfo {year}
  {2012})%
  \bibAnnoteFile{NoStop}{Harmon2012c}%
\bibitem{Grunewald2011}%
  \BibitemOpen
  \bibfield{author}{%
  \bibinfo {author} {\bibfnamefont{M.}~\bibnamefont{Gr\"unewald}}, \bibinfo
  {author} {\bibfnamefont{M.}~\bibnamefont{Wahler}}, \bibinfo {author}
  {\bibfnamefont{M.}~\bibnamefont{Michelfeit}}, \bibinfo {author}
  {\bibfnamefont{C.}~\bibnamefont{Gould}}, \bibinfo {author}
  {\bibfnamefont{R.}~\bibnamefont{Schmidt}}, \bibinfo {author}
  {\bibfnamefont{P.}~\bibnamefont{Graziosi}}, \bibinfo {author}
  {\bibfnamefont{A.}~\bibnamefont{Dediu}}, \bibinfo {author}
  {\bibfnamefont{F.}~\bibnamefont{W\"urthner}}, \bibinfo {author}
  {\bibfnamefont{G.}~\bibnamefont{Schmidt}},\ and\ \bibinfo {author}
  {\bibfnamefont{L.}~\bibnamefont{Molenkamp}},\ }%
  \bibfield{journal}{%
  \bibinfo {journal} {Phys. Rev. B}\ }%
  \textbf{\bibinfo {volume} {84}},\ \bibinfo {pages} {125208} (\bibinfo {year}
  {2011})%
  \bibAnnoteFile{NoStop}{Grunewald2011}%
\bibitem{Hellwig2007}%
  \BibitemOpen
  \bibfield{author}{%
  \bibinfo {author} {\bibfnamefont{O.}~\bibnamefont{Hellwig}}, \bibinfo
  {author} {\bibfnamefont{A.}~\bibnamefont{Berger}}, \bibinfo {author}
  {\bibfnamefont{J.~B.}\ \bibnamefont{Kortright}},\ and\ \bibinfo {author}
  {\bibfnamefont{E.~E.}\ \bibnamefont{Fullerton}},\ }%
  \bibfield{journal}{%
  \bibinfo {journal} {J. Magn. Magn. Mater.}\ }%
  \textbf{\bibinfo {volume} {319}},\ \bibinfo {pages} {13} (\bibinfo {year}
  {2007})%
  \bibAnnoteFile{NoStop}{Hellwig2007}%
\bibitem{Schellekens2011}%
  \BibitemOpen
  \bibfield{author}{%
  \bibinfo {author} {\bibfnamefont{A.~J.}\ \bibnamefont{Schellekens}}, \bibinfo
  {author} {\bibfnamefont{W.}~\bibnamefont{Wagemans}}, \bibinfo {author}
  {\bibfnamefont{S.~P.}\ \bibnamefont{Kersten}}, \bibinfo {author}
  {\bibfnamefont{P.~A.}\ \bibnamefont{Bobbert}},\ and\ \bibinfo {author}
  {\bibfnamefont{B.}~\bibnamefont{Koopmans}},\ }%
  \bibfield{journal}{%
  \bibinfo {journal} {Phys. Rev. B}\ }%
  \textbf{\bibinfo {volume} {84}},\ \bibinfo {pages} {075204} (\bibinfo {year}
  {2011})%
  \bibAnnoteFile{NoStop}{Schellekens2011}%
\bibitem{Shklovskii1984}%
  \BibitemOpen
  \bibfield{author}{%
  \bibinfo {author} {\bibfnamefont{B.~I.}\ \bibnamefont{Shklovskii}}\ and\
  \bibinfo {author} {\bibfnamefont{A.~L.}\ \bibnamefont{Efros}},\ }%
  \emph{\bibinfo {title} {Electronic Properties of Doped Semiconductors}}\
  (\bibinfo {publisher} {Springer},\ \bibinfo {address} {Heidelberg},\ \bibinfo
  {year} {1984})%
  \bibAnnoteFile{NoStop}{Shklovskii1984}%
\bibitem{Hu2007}%
  \BibitemOpen
  \bibfield{author}{%
  \bibinfo {author} {\bibfnamefont{B.}~\bibnamefont{Hu}}\ and\ \bibinfo
  {author} {\bibfnamefont{Y.}~\bibnamefont{Wu}},\ }%
  \bibfield{journal}{%
  \bibinfo {journal} {Nature Materials}\ }%
  \textbf{\bibinfo {volume} {6}},\ \bibinfo {pages} {985} (\bibinfo {year}
  {2007})%
  \bibAnnoteFile{NoStop}{Hu2007}%
\bibitem{Wagemans2011}%
  \BibitemOpen
  \bibfield{author}{%
  \bibinfo {author} {\bibfnamefont{W.}~\bibnamefont{Wagemans}}, \bibinfo
  {author} {\bibfnamefont{A.~J.}\ \bibnamefont{Schellekens}}, \bibinfo {author}
  {\bibfnamefont{M.}~\bibnamefont{Kemper}}, \bibinfo {author}
  {\bibfnamefont{F.~L.}\ \bibnamefont{Bloom}}, \bibinfo {author}
  {\bibfnamefont{P.~A.}\ \bibnamefont{Bobbert}},\ and\ \bibinfo {author}
  {\bibfnamefont{B.}~\bibnamefont{Koopmans}},\ }%
  \bibfield{journal}{%
  \bibinfo {journal} {Phys. Rev. Lett.}\ }%
  \textbf{\bibinfo {volume} {106}},\ \bibinfo {pages} {196802} (\bibinfo {year}
  {2011})%
  \bibAnnoteFile{NoStop}{Wagemans2011}%
\bibitem{Werner1977}%
  \BibitemOpen
  \bibfield{author}{%
  \bibinfo {author} {\bibfnamefont{H.~J.}\ \bibnamefont{Werner}}, \bibinfo
  {author} {\bibfnamefont{Z.}~\bibnamefont{Schulten}},\ and\ \bibinfo {author}
  {\bibfnamefont{K.}~\bibnamefont{Schulten}},\ }%
  \bibfield{journal}{%
  \bibinfo {journal} {The Journal of Chemical Physics}\ }%
  \textbf{\bibinfo {volume} {67}},\ \bibinfo {pages} {646} (\bibinfo {year}
  {1977})%
  \bibAnnoteFile{NoStop}{Werner1977}%
\bibitem{Rybicki2012}%
  \BibitemOpen
  \bibfield{author}{%
  \bibinfo {author} {\bibfnamefont{J.}~\bibnamefont{Rybicki}}, \bibinfo
  {author} {\bibfnamefont{R.}~\bibnamefont{Lin}}, \bibinfo {author}
  {\bibfnamefont{F.}~\bibnamefont{Wang}}, \bibinfo {author}
  {\bibfnamefont{M.}~\bibnamefont{Wohlgenannt}}, \bibinfo {author}
  {\bibfnamefont{C.}~\bibnamefont{He}}, \bibinfo {author}
  {\bibfnamefont{T.}~\bibnamefont{Sanders}},\ and\ \bibinfo {author}
  {\bibfnamefont{Y.}~\bibnamefont{Suzuki}},\ }%
  \bibfield{journal}{%
  \bibinfo {journal} {Phys. Rev. Lett.}\ }%
  \textbf{\bibinfo {volume} {109}},\ \bibinfo {pages} {076603} (\bibinfo {year}
  {2012})%
  \bibAnnoteFile{NoStop}{Rybicki2012}%
\bibitem{Timmel1998}%
  \BibitemOpen
  \bibfield{author}{%
  \bibinfo {author} {\bibfnamefont{C.~R.}\ \bibnamefont{Timmel}}, \bibinfo
  {author} {\bibfnamefont{U.}~\bibnamefont{Till}}, \bibinfo {author}
  {\bibfnamefont{B.}~\bibnamefont{Brocklehurst}}, \bibinfo {author}
  {\bibfnamefont{K.~A.}\ \bibnamefont{McLaughlin}},\ and\ \bibinfo {author}
  {\bibfnamefont{P.~J.}\ \bibnamefont{Hore}},\ }%
  \bibfield{journal}{%
  \bibinfo {journal} {Molecular Physics}\ }%
  \textbf{\bibinfo {volume} {95}},\ \bibinfo {pages} {71} (\bibinfo {year}
  {1998})%
  \bibAnnoteFile{NoStop}{Timmel1998}%
\bibitem{Gill1974}%
  \BibitemOpen
  \bibfield{author}{%
  \bibinfo {author} {\bibfnamefont{W.}~\bibnamefont{Gill}},\ }%
  \bibfield{journal}{%
  \bibinfo {journal} {Proceedings of the Fifth International Conference on
  Amorphous and Liquid Semiconductors}\ }%
  \textbf{\bibinfo {volume} {7222}},\ \bibinfo {pages} {901} (\bibinfo {year}
  {1974})%
  \bibAnnoteFile{NoStop}{Gill1974}%
\bibitem{Martens2000}%
  \BibitemOpen
  \bibfield{author}{%
  \bibinfo {author} {\bibfnamefont{H.~C.~F.}\ \bibnamefont{Martens}}, \bibinfo
  {author} {\bibfnamefont{P.~W.~M.}\ \bibnamefont{Blom}},\ and\ \bibinfo
  {author} {\bibfnamefont{H.~F.~M.}\ \bibnamefont{Schoo}},\ }%
  \bibfield{journal}{%
  \bibinfo {journal} {Phys. Rev. B}\ }%
  \textbf{\bibinfo {volume} {61}},\ \bibinfo {pages} {7489} (\bibinfo {year}
  {2000})%
  \bibAnnoteFile{NoStop}{Martens2000}%
\bibitem{Nguyen2010b}%
  \BibitemOpen
  \bibfield{author}{%
  \bibinfo {author} {\bibfnamefont{T.~D.}\ \bibnamefont{Nguyen}}, \bibinfo
  {author} {\bibfnamefont{B.~R.}\ \bibnamefont{Gautam}}, \bibinfo {author}
  {\bibfnamefont{E.}~\bibnamefont{Ehrenfreund}},\ and\ \bibinfo {author}
  {\bibfnamefont{Z.~V.}\ \bibnamefont{Vardeny}},\ }%
  \bibfield{journal}{%
  \bibinfo {journal} {Phys. Rev. Lett.}\ }%
  \textbf{\bibinfo {volume} {105}},\ \bibinfo {pages} {166804} (\bibinfo {year}
  {2010})%
  \bibAnnoteFile{NoStop}{Nguyen2010b}%
\bibitem{Cohen2009}%
  \BibitemOpen
  \bibfield{author}{%
  \bibinfo {author} {\bibfnamefont{A.}~\bibnamefont{Cohen}},\ }%
  \bibfield{journal}{%
  \bibinfo {journal} {J. Phys. Chem. A}\ }%
  \textbf{\bibinfo {volume} {113}},\ \bibinfo {pages} {11084} (\bibinfo {year}
  {2009})%
  \bibAnnoteFile{NoStop}{Cohen2009}%
\bibitem{Wang2008}%
  \BibitemOpen
  \bibfield{author}{%
  \bibinfo {author} {\bibfnamefont{F.~J.}\ \bibnamefont{Wang}}, \bibinfo
  {author} {\bibfnamefont{H.}~\bibnamefont{Bassler}},\ and\ \bibinfo {author}
  {\bibfnamefont{Z.~V.}\ \bibnamefont{Vardeny}},\ }%
  \bibfield{journal}{%
  \bibinfo {journal} {Phys. Rev. Lett.}\ }%
  \textbf{\bibinfo {volume} {101}},\ \bibinfo {pages} {236805} (\bibinfo {year}
  {2008})%
  \bibAnnoteFile{NoStop}{Wang2008}%
\end{thebibliography}
\end{document}